\documentclass[12pt]{iopart}
\usepackage{iopams}  

\usepackage{setstack}
\usepackage{cite}
\usepackage{graphicx}
\usepackage{hyperref}
\usepackage[usenames,dvipsnames]{xcolor}
\hypersetup{colorlinks=true, linkcolor=BrickRed, urlcolor=blue!50!black, citecolor=blue!50!black}

\begin{document}

\title{Tagged-particle motion in quasi-confined colloidal hard-sphere liquids}

\author{Lukas Schrack$^1$, Charlotte F. Petersen$^{1,2}$, Michele Caraglio$^1$, Gerhard Jung$^1$ and Thomas Franosch$^1$}
\address{$^1$ Institut f\"ur Theoretische Physik, Universit\"at Innsbruck, Technikerstraße 21A, 6020 Innsbruck, Austria}
\address{$^2$ Centre for Theoretical and Computational Molecular Science, Australian Institute for Bioengineering and Nanotechnology, The University of Queensland, Brisbane, QLD 4072, Australia}

\ead{lukas.schrack@uibk.ac.at, charlotte.petersen@uq.edu.au, thomas.franosch@uibk.ac.at}
\vspace{10pt}
\begin{indented}
\item[]\today
\end{indented}

\begin{abstract}
We investigate the tagged-particle motion in a strongly interacting quasi-confined liquid using periodic boundary conditions along the confining direction. Within a mode-coupling theory of the glass transition (MCT) we calculate the self-nonergodicity parameters and the self-intermediate scattering function and compare them with event-driven molecular dynamics simulations. We observe non-monotonic behavior for the in-plane mean-square displacement and further correlation functions which refer to higher mode indices encoding information about the perpendicular motion. The in-plane velocity-autocorrelation function reveals persistent anti-correlations with a negative algebraic power-law decay $t^{-2}$ at all packing fractions.
\end{abstract}

\vspace{2pc}

\submitto{\JSTAT}
%\maketitle

\section{Introduction}
Systems displaying glassy dynamics are prevalent in nature ranging from granular matter to colloids to biological applications. For example, experiments demonstrate that bacterial cytoplasms become glassy in the case of a suppressed metabolism~\cite{Bradley:Cell:2014, Nishizawa:SciRep:2017}. Recently, the importance of glassy physics has also been revealed in numerous studies on active particles~\cite{Berthier:JCP:2019, Janssen:JoP:2019, Lozano:NatureMaterial:2019, Szamel:JCP:2019}. Furthermore, significant progress on the slow structural relaxation has been achieved by using machine learning techniques~\cite{Bapst:NaturePhysics:2020} and novel optimized algorithms in computer simulations~\cite{Ninarello:PRX:2017}. Yet, the nature of the glass transition is still heavily debated~\cite{Berthier:Dynamical_Heterogeneities} and remains one of the grand challenges of condensed matter physics~\cite{Anderson:Science:1995}. The mode-coupling theory of the glass transition (MCT) provides an entirely first-principle description to elucidate this complex behavior~\cite{Goetze:Complex_Dynamics, Janssen:FiP:2018}. Other approaches are dynamical mean-field theory~\cite{Maimbourg:PRL:2016, Manacorda:JCP:2020} and self-consistent generalized Langevin dynamics~\cite{Yeomans-Reyna:PRE_64:2001, Yeomans-Reyna:PRE_76:2007}. Only recently, mode-coupling-like approaches have been successfully extended to systems of active Brownian particles~\cite{Farage:PRE:2015, Liluashvili:PRE:2017, Szamel:JCP:2019} and self-propelled particles~\cite{Szamel:PRE:2016, Feng:SoftMatter:2017, Nandi:SoftMatter:2017}.

Indisputably, confined systems are of particular interest due to their presence in nature and industrial applications. Since local structure plays a predominant role for the slowing down of the dynamics upon approaching the glass transition, confinement strongly affects both structure and dynamics. 
This complicated interplay of confinement with the cage effect has been extensively investigated within the slit geometry in both experiments~\cite{Nugent:PRL:2007, Nygard:PRL:2012, Nygard:PRL:2016} and hard-sphere simulations~\cite{Torres:PRL:2000, Mittal:PRL:2006, Mandal:NatComm:2014}.

MCT has been successfully extended to these inhomogeneous confined liquids using symmetry-adapted modes~\cite{Lang:PRL:2010, Lang:PRE:2012}. Other inhomogeneous MCT approaches incorporate (infinitesimal) local external fields~\cite{Biroli:PRL:2006} or periodic potential landscapes~\cite{Nandi:PRE:2011}.

Although its relative simplicity, which makes it one of the most widely studied confined systems, the slit geometry already incorporates a non-trivial interplay between local packing (caging) and the inhomogeneous density profile induced by the walls (layering). However, in a slit geometry it is not possible to unravel the effects of confinement and those simply due to layering. To circumvent this problem, quasi-confined liquids are considered~\cite{Petersen:JStatMech:2019, Schrack:JStatMech:2020}, where periodic boundary conditions in the short confining direction are introduced. In this set-up, the liquid becomes translationally invariant both in the unconfined as well as in the confined direction, with a uniform density profile. Yet, there is still an impact from the confinement on the particles once the characteristic confinement length becomes comparable to the particle size.

The goal of the present work is to advance the analysis of quasi-confined liquids within the framework of MCT and with event-driven molecular dynamics simulations. The collective behavior has been discussed in detail recently~\cite{Schrack:JStatMech:2020}. Even in absence of layering a reentrance behavior is observed in the non-equilibrium state diagram, motivating us to further investigate these liquids, in particular with regard to the self-dynamics. In principle, the tagged-particle motion is directly accessible in experiments using neutron scattering, dynamic light scattering or single particle tracking. Due to enhanced statistics it is also favorable to extract self-correlation functions from computer simulations, which is of particular interest for this work.

The tagged-particle MCT equations for the general class of layered fluids have been derived earlier for Newtonian dynamics~\cite{Lang:PRE_89:2014} and have been extended to Brownian dynamics more recently~\cite{Schrack:PhilMag:2020}. As for the collective dynamics, the only required input to these MCT equations are the static quantities of quasi-confined liquids~\cite{Petersen:JStatMech:2019}.

If the self-intermediate scattering function (SISF) is known, all moments of the displacement such as the mean-square displacement (MSD) can be extracted. The long-time behavior of the MSD provides the diffusion coefficient. Generally, the self-diffusion coefficient is linked to the velocity-autocorrelation function (VACF) by a Green-Kubo relation. It is well established that the VACF displays an algebraic power-law decay $t^{-3/2}$ in three-dimensional systems for underlying Newtonian dynamics~\cite{Alder:PRL:1967, Alder:PRA:1970}. Similar anti-correlations with a tail $t^{-5/2}$ are a general feature of strongly interacting systems without momentum conservation due to particle conservation. Only recently, these tails have been corroborated using Brownian dynamics simulations and MCT~\cite{Mandal:PRL:2019}.

We start with a short summary of the relevant MCT equations for the dynamics of a tracer particle, which resembles the collective equations~\cite{Schrack:JStatMech:2020}. In the remaining part of \sref{sec:theoretical_description} we present the relevant equations for further tagged-particle quantities, in particular the MSD and the VACF. Additionally, the long-time behavior is discussed. The results within the framework of MCT are qualitatively compared with computer simulations in \sref{sec:results}. We conclude in \sref{sec:conclusion} with a critical evaluation of our findings.

\section{Mode-coupling theory}\label{sec:theoretical_description}
\subsection{Tagged-particle motion}
In this paper we investigate the dynamics of a tracer (tagged) particle with diameter $\sigma_s$ and bare diffusion coefficient $D_0^{(s)}$ suspended in a 3D quasi-confined colloidal liquid of identical hard spheres with diameter $\sigma$ and bare diffusion coefficient $D_0$. All particles evolve according to Brownian dynamics, where hydrodynamic interactions are ignored and are restricted to a quasi-confined environment by applying periodic boundary conditions along the confining direction, which we refer to as the $z$-direction. Thus, the transversal coordinates are folded back to $-L/2\le z\le L/2$. The in-plane coordinates are denoted by $\vec{r}=(x,y)$ and are infinitely extended in the thermodynamic limit, $N\to\infty$, $A\to\infty$, where $N$ is the number of particles and $A$ is the area in the dimensions perpendicular to the confining direction. We introduce the area density $n_0=N/A$ and the volume density $n=n_0/L$ related to the packing fraction $\varphi=n\pi\sigma^3/6$ of the system. All coordinates and quantities of the tracer particle are labeled with a superscript $s$.

In this section, we convert the relevant MCT equations from the more general class of layered colloidal liquids~\cite{Schrack:PhilMag:2020} to quasi-confined fluids. For this purpose we basically follow reference~\cite{Schrack:JStatMech:2020} in which the corresponding collective equations have been derived. Since most of the steps of the derivation are identical if collective quantities are replaced by the corresponding single-particle quantities, we just provide a short summary of the main equations, focusing on the relevant differences to the collective case.

The most relevant correlation function is the SISF, also called the incoherent intermediate scattering function,
\begin{eqnarray}\label{eq:SISF}
 S_{\mu}^{(s)}(q,t)=\langle \rho_\mu^{(s)}(\vec{q},t)^*\rho_\mu^{(s)}(\vec{q})\rangle,
\end{eqnarray}
given in terms of the fluctuating density modes 
\begin{eqnarray}
 \rho_\mu^{(s)}(\vec{q})=\exp{\left[\rmi Q_\mu z_s\right]}e^{\rmi \vec{q}\cdot\vec{r}_s},
\end{eqnarray}
of the tagged particle. The initial value $S_\mu^{(s)}(q,t=0)=1$ is independent of $q$ and $\mu$. The two-dimensional wavevector $\vec{q}=(q_x,q_y)$ in the lateral direction becomes continuous in the thermodynamic limit, whereas the wavenumber $Q_{\mu}=2\pi\mu/L,\mu\in\mathbb{Z}$ is discrete. For $L\to\infty$ rotational invariance with the 3D wavevector $\vec{k}=(\vec{q},Q_\mu)$ is restored and the bulk limit is reached as it has been demonstrated for the general class of layered fluids~\cite{Lang:PRE_90:2014}. Due to periodic boundary conditions, the accessible $z$-coordinate positions do not differ between the tracer and the host-liquid particles, even if the tracer particle varies in size. This is in contrast to the case of a tracer particle confined between two parallel walls, where the effective confinement length $L_s$ perpendicular to the walls  for the tracer particle does not have to coincide with the accessible slit width $L$ for the host-liquid particles~\cite{Lang:PRE_89:2014}. Therefore, for the tagged particle within the slit geometry an additional plane wave basis $\exp{(-\rmi Q_\mu^{(s)}z)}$, with $Q_{\mu}^{(s)}=2\pi\mu/L_s,\mu\in\mathbb{Z}$ is required. Quasi-confinement, however, simplifies the situation and no additional plane wave basis is necessary.

The exact equations of motion (e.o.m.)\ for $S_{\mu}^{(s)}(q,t)$ can be derived using the Zwanzig projection operator formalism~\cite{Goetze:Complex_Dynamics} introducing an effective self-memory kernel $M_\mu^{(s)}(q,t)$ analogous to the collective case~\cite{Schrack:JStatMech:2020}
\begin{eqnarray}\label{eq:eom_tagged}
 \fl\dot{S}_\mu^{(s)}(q,t)+&D_\mu^{(s)}(q)S_\mu^{(s)}(q,t) + D_\mu^{(s)}(q)\int_0^t M_\mu^{(s)}(q,t-t^\prime)\dot{S}_\mu^{(s)}(q,t^\prime)\mathrm{d} t^\prime = 0,
\end{eqnarray}
with initial decay $D_\mu^{(s)}(q)=D_0^{(s)}\left(q^2+Q_\mu^{2}\right)$ related to the tagged-particle bare diffusion coefficient $D_0^{(s)}$. Due to the quasi-confined geometry the relaxation naturally splits into channels parallel and perpendicular to the confining direction. Following the same procedure as for the collective case, the effective kernel is related to the matrix-valued irreducible memory kernel $\boldsymbol{\mathcal{M}}_\mu^{(s)}(q,t)$ with diagonal components $\mathcal{M}_\mu^{\parallel\parallel,(s)}(q,t)$ and $\mathcal{M}_\mu^{\perp\perp,(s)}(q,t)$ and the symmetric off-diagonal element $\mathcal{M}_\mu^{\parallel\perp,(s)}(q,t)$ by the relation
\begin{eqnarray}\label{eq:effective_memory_time_tagged}
 D_{\mu}^{(s)}(q)  M_{\mu}^{(s)}(q,t) &+ D_0^{(s)2} \int_0^t M_{\mu}^{(s)}(q,t-t^\prime)\alpha_{\mu}^{(s)}(q,t^\prime)\mathrm{d}t^\prime =  D_0^{(s)} \beta_{\mu}^{(s)}(q,t) \nonumber\\
 &+ D_0^{(s)2}\int_0^t \mathcal{M}_{\mu}^{\parallel\parallel,(s)}(q,t-t^\prime)
  \mathcal{M}_{\mu}^{\perp\perp,(s)}(q,t^\prime)\mathrm{d}t^\prime \nonumber \\ &- D_0^{(s)2}\int_0^t \mathcal{M}_{\mu}^{\parallel\perp,(s)}(q,t-t^\prime)\mathcal{M}_{\mu}^{\parallel\perp,(s)}(q,t^\prime)\mathrm{d}t^\prime.
\end{eqnarray}
Therein, the coefficients
\begin{eqnarray}
\fl\alpha_{\mu}^{(s)}(q,t) = Q_{\mu}^{2}\mathcal{M}_{\mu}^{\parallel\parallel,(s)}(q,t) + q^2\mathcal{M}_{\mu}^{\perp\perp,(s)}(q,t) - 2 q Q_{\mu}\mathcal{M}_{\mu}^{\parallel\perp,(s)}(q,t),
\end{eqnarray}
and
\begin{eqnarray}
\fl\beta_{\mu}^{(s)}(q,t) = \frac{q^2}{q^2+Q_{\mu}^{2}} \mathcal{M}_{\mu}^{\parallel\parallel,(s)}(q,t) + \frac{Q_{\mu}^{2}}{q^2 + Q_{\mu}^{2}}\mathcal{M}_{\mu}^{\perp\perp,(s)}(q,t) + \frac{2q Q_{\mu}}{q^2+Q_{\mu}^{2}}\mathcal{M}_{\mu}^{\parallel\perp,(s)}(q,t),
\end{eqnarray}
are completely determined by the matrix entries of the irreducible memory kernel $\boldsymbol{\mathcal{M}}_\mu^{(s)}(q,t)$.

Finally, MCT approximates the irreducible memory kernel as a bilinear functional of the ISF and the SISF, where the coupling between different modes and wavenumbers is described by vertices. These are entirely determined by the direct correlation function $c_\mu^{(s)}(q)$ of the tagged particle, which is connected to the static structure factor $S_\mu^{(s)}(q)$ by a generalized Ornstein-Zernike relation~\cite{Petersen:JStatMech:2019}. Numerical efficiency can be remarkably enhanced by introducing bipolar coordinates~\cite{Schrack:JStatMech:2020}
\begin{eqnarray}
  \fl\mathcal{M}_\mu^{\alpha\beta,(s)}(q,t) = \frac{n}{4L^3\pi^2}\int_0^\infty\int_{|q-q_1|}^{q+q_1} & \sum_{\substack{\mu_1=-\infty\cr \mu_2=\mu-\mu_1}}^{\infty} S_{\mu_1}(q_1,t)  S_{\mu_2}^{(s)}(q_2,t)\frac{4q_1q_2} {\sqrt{4q^2q_1^2-(q^2+q_1^2-q_2^2)^2}}  \nonumber \\
 &\times\left[b^\alpha\left(\frac{q^2+q_1^2-q_2^2}{2q},Q_{\mu_1}\right)c_{\mu_1}^{(s)}(q_1)\right] \nonumber\\
 & \times\left[b^\beta\left(\frac{q^2+q_1^2-q_2^2}{2q},Q_{\mu_1}\right)c_{\mu_1}^{(s)}(q_1)\right]\mathrm{d} q_1\mathrm{d} q_2,
\end{eqnarray}
where $b^\alpha(x,z)=x\delta_{\alpha \parallel}+z\delta_{\alpha \perp}$ is the selector with respect to the channel indices $\alpha, \beta \in \{\parallel,\perp\}$. The indices $\mu_1$ and $\mu_2$ run over all integer values satisfying the selection rule $\mu=\mu_1+\mu_2$, which is valid due to translational invariance along the confining direction. Since not only the SISF, but also the ISF enter the equations and all tagged-particle quantities within MCT are derived quantities, the collective dynamics have to be solved first. The algorithm for the tagged-particle dynamics is equivalent to the collective case. Solving the coupled equations for the effective (self-)memory kernel and the (S)ISF is not straightforward. A detailed description of the algorithm using a decimation scheme can be found in references~\cite{Schrack:JStatMech:2020, Gruber:PRE:2016}.

To distinguish ergodic liquid states from arrested localized states it is useful to introduce the nonergodicity parameter or glass-form factor as the long-time limit of the SISF,
\begin{eqnarray}
F_\mu^{(s)}(q) := \lim_{t\to\infty} S_\mu^{(s)}(q,t).
\end{eqnarray}
Liquid states then correspond to vanishing nonergodicity parameters, whereas $F_\mu^{(s)}(q)$ remains finite for glassy states. The nonergodicity parameters are directly accessible in simulations or experiments, and within the framework of MCT $F_\mu^{(s)}(q)$ can be determined without solving the full time-dependent equations explicitly~\cite{Schrack:JStatMech:2020,Lang:PRE_89:2014}.

\subsection{Long-wavelength limit}
We now elaborate only the lateral dynamics of the quasi-confined liquid from a ``top-view'' perspective and investigate the mean-square displacement (MSD) and the velocity-autocorrelation function (VACF). Even though no confinement is present in the parallel direction, the dynamics of this two-dimensional projection are still sensitive to the confining length. To derive explicit equations for the MSD and the VACF we investigate the tagged-particle dynamics in the long-wavelength limit. The small $q$ series of the SISF 
\begin{eqnarray}\label{eq:SISF_small_q}
S_\mu^{(s)}(q,t)=P_\mu(t)-q^2 \delta r_\mu^{2} (t)/4 + \mathcal{O}(q^4),
\end{eqnarray}
together with the out-of-plane self-intermediate scattering function
\begin{eqnarray}\label{eq:P_mu}
P_\mu(t)=\left\langle\exp{\left[-\rmi Q_\mu\Big(z(t)-z(0)\Big)\right]}\right\rangle,
\end{eqnarray}
can be written in terms of a generalized MSD 
\begin{eqnarray}
\delta r_\mu^2(t) := \left\langle\exp\left[-\rmi Q_\mu\Big(z(t)-z(0)\Big)\right]\left[\vec{r}(t)-\vec{r}(0)\right]^2\right\rangle.
\end{eqnarray}
$P_\mu(t)$ does not depend on the in-plane wavevector $\vec{q}$, and probes the dynamics of the tracer only along the confinement direction.
Inserting the expansion for $S_0^{(s)}(q,t)$ with $P_0(t)=1$ in the e.o.m.\ for the tagged-particle correlator, equation~\eref{eq:eom_tagged}, we obtain for the MSD
\begin{eqnarray}
 \delta r^2_{0}(t)+D_0^{(s)}\beta^2\int_0^t \zeta^{(s)}_{0}(t-t^\prime)\delta r^2_{0}(t^\prime)\mathrm{d} t^\prime = 4 D_0^{(s)} t,
\end{eqnarray}
reminiscent of the corresponding e.o.m.\ in 2D systems~\cite{Bayer:PRE:2007}. However, in the case of quasi-confinement, a second equation for the long-wavelength limit of the effective self-memory kernel $\beta^2\zeta_0^{(s)}(t)=\lim_{q\to 0}q^2M_0^{(s)}(q,t)$ is necessary. The irreducible memory kernel $\boldsymbol{\mathcal{M}}_\mu^{(s)}(q,t)$ for arbitrary mode index $\mu$ decouples in the long-wavelength limit, $\lim_{q\to 0}\mathcal{M}_\mu^{\alpha\beta,(s)}(q,t)=\mathcal{M}_\mu^{\alpha,(s)}(t)\delta_{\alpha\beta}$. The non-vanishing diagonal components are then given by
\begin{eqnarray}\label{eq:memory_parallel}
\mathcal{M}_\mu^{\parallel,(s)} (t) = \frac{n}{4\pi L^3} \int & k^3 \mathrm{d} k\,\sum_{\mu_1}\left(c_{\mu_1}^{(s)}(k)\right)^2 S_{\mu_1}(k,t) S_{\mu-\mu_1}^{(s)}(k,t),
\end{eqnarray}
and
\begin{eqnarray}\label{eq:memory_perp}
\mathcal{M}_\mu^{\perp,(s)} (t) = \frac{n}{2\pi L^3} \int & k\mathrm{d} k\, \sum_{\mu_1}\left(Q_{\mu_1}c_{\mu_1}^{(s)}(k)\right)^2 S_{\mu_1}(k,t) S_{\mu-\mu_1}^{(s)}(k,t),
\end{eqnarray}
enabling us to determine the equation for the effective memory kernel $\beta^2\zeta_0^{(s)}(t)$ in the long-wavelength limit
\begin{eqnarray}\label{eq:effective_memory}
D_0^{(s)} \beta^2\zeta_0^{(s)}(t) &+ D_0^{(s)2} \beta^2 \int_0^t \zeta_0^{(s)}(t-t^\prime)\mathcal{M}_0^{\perp,(s)}(t^\prime)\mathrm{d} t^\prime = D_0^{(s)} \mathcal{M}_0^{\parallel,(s)} (t) \nonumber\\ 
&+ D_0^{(s)2} \int_0^t \mathcal{M}_0^{\parallel,(s)}(t-t^\prime)\mathcal{M}_0^{\perp,(s)}(t^\prime)\mathrm{d} t^\prime.
\end{eqnarray}
The equation for the effective memory kernel is not restricted to $\mu=0$ relevant for the MSD, but can also be extended to arbitrary mode indices 
\begin{eqnarray}
D_0^{(s)} Q_\mu^{2} M_\mu^{(s)}(t) &+ D_0^{(s)2}\int_0^t Q_\mu^{2} M_\mu^{(s)}(t-t^\prime)\mathcal{M}_\mu^{\parallel,(s)}(t^\prime)\mathrm{d} t^\prime 
= D_0^{(s)} \mathcal{M}_\mu^{\perp,(s)}(t) \nonumber\\ 
&+ D_0^{(s)2}\int_0^t \mathcal{M}_\mu^{\parallel,(s)}(t-t^\prime)\mathcal{M}_\mu^{\perp,(s)}(t^\prime)\mathrm{d} t^\prime, \quad\mu\neq 0,
\end{eqnarray}
with the long-wavelength limit $M_\mu^{(s)}(t)=\lim_{q\to 0}M_\mu^{(s)}(q,t)$ for higher order effective self-memory kernels.

From the expansion for $S_\mu^{(s)}(q,t)$, equation~\eref{eq:SISF_small_q}, we get
\begin{eqnarray}
\dot{P}_\mu(t)&+D_0^{(s)}Q_\mu^2 P_\mu(t) +D_0^{(s)}\int_0^t Q_\mu^2 M_\mu(t-t^\prime)\dot{P}_\mu(t^\prime)\mathrm{d} t^\prime=0,
\end{eqnarray}
as the e.o.m.\ for $P_\mu(t)$. In contrast to the MSD, which only considers the dynamics in the parallel direction, it takes into account the dynamics perpendicular to the confining direction. The mathematical structure of this integro-differential equation is reminiscent of the e.o.m.\ for the tagged-particle correlator, equation~\eref{eq:eom_tagged}, and therefore it is straightforward to solve it numerically.

The velocity-autocorrelation function (VACF) $Z_0(t)$ of a quasi-confined colloidal liquid can formally be connected to the MSD by the relation
\begin{eqnarray}\label{eq:VACF}
Z_0(t) := \frac{1}{4}\frac{\mathrm{d}^2}{\mathrm{d} t^2}\delta r^2_0(t).
\end{eqnarray}
Therefore, the in-plane VACF can in principle be determined by taking derivatives of the MSD. Yet, due to discretization of the wavenumber grids in MCT, long-wavelength phenomena cannot be properly resolved. An exact e.o.m.\ can be readily derived using the Zwanzig projection operator formalism with the same irreducible effective memory kernel $\beta^2\zeta_0^{(s)}(t)$ as already introduced above in the e.o.m.\ for the MSD
\begin{eqnarray}\label{eq:eom_VACF}
 Z_0(t)+ D_0^{(s)}\beta^2\int_0^t \zeta_0^{(s)}(t-t^\prime)Z_0(t^\prime)\mathrm{d} t^\prime = -D_0^{(s)2}\beta^2\zeta_0^{(s)}(t).
\end{eqnarray}

\subsection{Long-time behavior}

We elaborate the long-time behavior of the VACF within the framework of MCT explicitly.  For long times, the integrals for the irreducible memory kernel, equations~\eref{eq:memory_parallel} and \eref{eq:memory_perp}, are dominated by small wavenumbers. In the long-wavelength limit, $\mu=0$ is the only mode which becomes slow and contributes to the sum. Thus, the perpendicular component vanishes, $\mathcal{M}_0^{\perp,(s)}(t)\simeq 0$. Correspondingly, for long times the equation for the effective memory kernel, equation~\eref{eq:effective_memory}, reduces to $D_0^{(s)}\beta^2\zeta_0^{(s)}(t) = D_0^{(s)} \mathcal{M}_0^{\parallel,(s)}(t)$. Approximating the ISF and SISF for $\mu=0$ by collective diffusive motion, $S_0(q,t)\approx S_0(0)\exp\left(-D_0 q^2 t /S_0(0)\right)$ and self-diffusion $S_0^{(s)}(q,t)\approx \exp\left(-D_{0}^{L,(s)}q^2t\right)$, the long-time behavior of the VACF follows from the equation of motion, equation~\eref{eq:eom_VACF}, using Tauber's theorem $Z(t)\simeq-(D_0^{L,(s)})^2 \beta^2\zeta_0^{(s)}(t)$. The explicit representation of the long-time asymptote reads
\begin{eqnarray}\label{eq:VACF_LTT}
Z_0(t) \simeq -\left(D_0^{L,(s)}\right)^2 \frac{n}{4\pi L^3} \frac{\left(c_{0}^{(s)}(0)\right)^2S_{0}(0)}{2\left(D_0^{L,(s)}+D_0/S_0(0)\right)^2} t^{-2}.
\end{eqnarray}
This power-law tail $\sim -t^{-2}$ is similar to the persistent anti-correlations found in bulk systems~\cite{Mandal:PRL:2019}. Due to the confinement, long-wavelength phenomena are only important along the lateral direction and we identify for long times the expected power-law for a two-dimensional system with a different prefactor.

In principle, autocorrelation functions similar to the VACF can also be defined for higher order modes, $\mu \ge 1$, characterizing the out-of-plane dynamics with the same approach as for the MSD. However, in this work we focus on the long-time behavior of the VACF, since no long-time tails are expected for $\mu\ne 0$.

\section{Simulations and numerical results}\label{sec:results}
After describing the set-up for the computer simulations, we analyze the tagged-particle quantities (self-nonergodicity parameters, SISF, MSD and VACF) from a polydisperse hard-sphere fluid and compare them with our results from MCT. The long-time behavior of the VACF is solely analyzed within the framework of MCT. The tracer particle is identical to the bath particles with the same diameter $\sigma$ and bare diffusion coefficient $D_0$. Accordingly, it becomes particularly simple to improve statistics in the simulation results. As input for the MCT calculations we use the static structure factors $S_\mu(q)$ and direct correlation functions $c_\mu(q)$ elaborated within liquid state theory using the Percus-Yevick (PY) closure relation. A detailed comparison between PY and monodisperse simulations is done in reference~\cite{Petersen:JStatMech:2019} and a short assessment of the differences between PY and polydisperse simulation data has been performed recently in connection with the discussion of the collective dynamics in quasi-confined systems~\cite{Schrack:JStatMech:2020}.

\subsection{Simulations}
We perform event-driven molecular dynamics simulations for a 3D hard-sphere system, with periodic boundary conditions in all directions. The quasi-confinement is realized by setting the length of the $z$-direction of the simulation box much smaller than the other two directions. The extension of the unconfined directions is $25\sigma$, whereas the confining length, $L$, has only a few particle diameters. Although the simulated particles are driven by Newtonian dynamics, we expect that the qualitative behavior, in particular for long times, will be comparable to MCT results for colloidal liquids undergoing Brownian motion~\cite{Gleim:PRL:1998, Franosch:JNCS:1998, Hunter:RepProgPhys:2012, Pusey:PhilTransRoyal:2009}. To prevent crystallization polydisperse particles are used in the simulations. The particle diameters are given by an inverse-occupied volume distribution~\cite{Ninarello:PRX:2017} with a standard deviation of $0.117\sigma$, setting the polydispersity to $11.7\%$. It must be ensured that particles are initialized without overlap, which we do here by initializing particles in a larger volume, and compressing the system with standard algorithms~\cite{Woodcock:Annals:1981, Li:EPL:2008} to reach the desired packing fraction and system size~\cite{Schrack:JStatMech:2020}. The time scale of the simulation is determined by the thermal energy $k_BT$ and the particle mass $m$, where we define $t_0=\sqrt{m\sigma^2/k_BT}$. The SISF can directly be calculated from the density-mode representation, equation~\eref{eq:SISF}. In general, statistics for incoherent quantities can be significantly enhanced compared to collective ones by averaging over all particles. 

As we have discussed previously~\cite{Schrack:JStatMech:2020}, it is not practical to reach a fully equilibrated glassy fluid of polydisperse hard spheres in quasi-confinement, because as in the slit geometry~\cite{Jung:PRR:2020}, the particles will eventually demix and then crystallize. However, this process occurs on long timescales, outside what is experimentally relevant, and only accessible in simulations when relying on advanced, unphysical dynamical algorithms \cite{Berthier:PRL:2016, Ninarello:PRX:2017}. Therefore, to study glassy dynamics here, we use Newtonian dynamics only, restrict ourselves to relatively short simulation times, and carefully check the results for any effects of aging. Our data is taken over a time $10^5 t_0$, after a sample preparation time $10^5 t_0$. We have also repeated all simulations using a longer sample preparation time of $10^6 t_0$, and found this does not change the qualitative behavior we will discuss in this section, and therefore the conclusions drawn are not affected by aging.

The simulation data for the SISF can be approximately described by the phenomenological Kohlrausch-William-Watts (KWW) stretched exponential~\cite{Williams:Faraday:1970}
\begin{eqnarray}
\mbox{KWW fit: } S_\mu^{(s)}(q,t)=F_\mu^{(s)}(q)\exp\left\{-\left[t/\tau_\mu^{(s)}(q)\right]^{\beta_\mu^{(s)}(q)}\right\}.
\end{eqnarray}
The data is fit in the time range $t\in(10 t_0,10^5 t_0)$ and the incoherent quantities, namely the Kohlrausch exponent $\beta_\mu^{(s)}(q)$, the relaxation time $\tau_\mu^{(s)}(q)$ and the self-nonergodicity parameter $F_\mu^{(s)}(q)$, can be extracted.

\subsection{Self-nonergodicity parameter}
Unless otherwise stated, the numerical results are obtained within MCT working on an equidistant wavenumber grid $q=\hat{q}\Delta q + q_0$ with grid points $\hat{q}=0,\dots,N_q -1$ parallel to the confinement. As input parameters $q_0\sigma=0.1212$, $\Delta q \sigma = 0.404$ and $N_q=100$ are used. The discrete mode index associated with the confining direction is limited to $|\mu|\le 15$. We assume that the tagged particle is of the same species as the bath particles, therefore the identical direct correlation functions as in the collective case enter the MCT equations.

\begin{figure}[htp]
\centering
\includegraphics{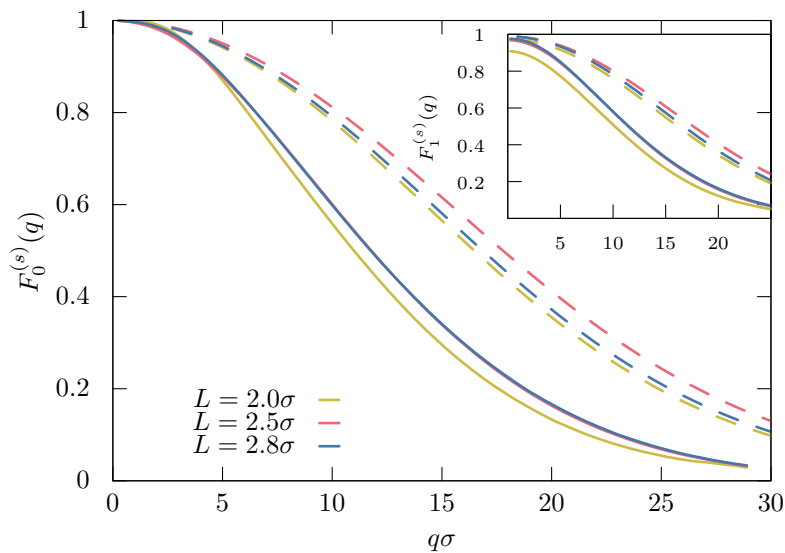}
\caption{Incoherent nonergodicity parameters $F_0^{(s)}(q)$ and $F_1^{(s)}(q)$ (inset) for different values of $L$ at packing fraction $\varphi=0.53$ for MCT (dashed lines) and $\varphi=0.59$ for simulations (solid lines). The red and blue lines for the simulations are practically on top of each other.}
\label{fig:nonergodicity}
\end{figure}

We start by analyzing the non-monotonic behavior of the nonergodicity parameters depending on the confinement length. We choose three different lengths $L=2.0\sigma, 2.5\sigma, 2.8\sigma$ where the effects are most pronounced. The non-monotonic behavior coincides qualitatively with the collective case~\cite{Schrack:JStatMech:2020}, therefore we believe that discussing more different values does not contribute to a deeper understanding. Since it is known that MCT underestimates the critical packing fraction not only in the bulk, but also for confined systems~\cite{Mandal:NatComm:2014, Mandal:SoftMatter:2017}, MCT results for $\varphi=0.53$ (\fref{fig:nonergodicity} (dashed lines)) are compared to simulations with a higher packing fraction $\varphi=0.59$ (\fref{fig:nonergodicity} (solid lines)), therefore the dynamics are glassy in both cases. In principle, this discrepancy in the packing fraction can be reduced with modified MCT approaches~\cite{Luo:JCP:2020}, but it does not affect the validity of our conclusions in this work. Referring to the nonequilibrium-state diagram of quasi-confined liquids~\cite{Schrack:JStatMech:2020}, MCT results are presented for $\varphi=0.53$ in contrast to the remainder of this manuscript to ensure that the behavior is glassy for all investigated confinement lengths.

Within MCT, the bell-shaped curves manifest non-monotonic behavior for all wavenumbers. The incommensurate value $L=2.5\sigma$ shows a higher structural arrest compared to the more commensurate values $L=2.0\sigma$ and $L=2.8\sigma$. The confinement length is referred as commensurate if the ratio $L/\sigma$ is close to an integer and as incommensurate if it is close to a half-integer. The non-monotonic behavior reflects the collective behavior in quasi-confinement~\cite{Schrack:JStatMech:2020} and is also quite similar to the incoherent nonergodicity parameter within the slit geometry~\cite{Mandal:SoftMatter:2017}, although the non-monotonic behavior is less noticeable due to missing layering.

In contrast, for the simulations non-monotonic behavior is not obvious. Indeed, the structural arrest for $L=2.5\sigma$ is higher compared to $L=2.0\sigma$, but it is almost indistinguishable from $L=2.8\sigma$, where we expect a significant drop in the structural relaxation. Using the von Schweidler law~\cite{Goetze:Complex_Dynamics} instead of KWW fits, a marginal difference can be observed. Yet, already in the slit case the non-monotonic behavior of the nonergodicity parameters is less pronounced for the simulations compared to the MCT results~\cite{Mandal:SoftMatter:2017}. Since effects are considerably smaller in quasi-confinement anyway, the overlap of the curves for $L=2.8\sigma$ and $L=2.8\sigma$ is not surprising. This behavior indicates that the height of the plateau value of the SISF does not show a pronounced non-monotonic behavior, which is most likely due to the inclusion of polydispersity, in contrast to the MCT predictions and simulation data for the collective motion~\cite{Schrack:JStatMech:2020}. It has been shown for bulk systems~\cite{Weysser:PRE:2010} that the incoherent nonergodicity parameter is less sensitive to effects of different particle sizes of the constituents compared to the collective case. Therefore, we conclude that also effects of local order between commensurate and incommensurate packing have a minor influence on the incoherent nonergodicity parameter.

For both MCT and simulations the lowest order mode $F_0^{(s)}(q)$ reaches unity in the long-wavelength limit reflecting particle conservation in the system. For higher order modes, $F_\mu^{(s)}(q\to 0)$ corresponds to $P_\mu(t\to\infty)$ according to equation~\eref{eq:SISF_small_q}, with a value different from unity in agreement with equation~\eref{eq:P_mu}. $P_\mu(t)$ is only sensitive along the confining direction, and for large confinement length the nonergodicity parameter should approach the bulk limit. Then, it only depends on the magnitude of the 3D wave vector $k=\sqrt{q^2+Q_\mu^{2}}$ and rotational invariance is restored, $F_\mu^{(s)}(q)\to F^{(s)}(k)$. The drop between $F_0^{(s)}(q\to 0)$ and $F_1^{(s)}(q\to 0)$ can be observed in \fref{fig:nonergodicity} for simulations, whereas for MCT it becomes only visible for higher order modes, $\mu\ge 2$, not included in the figure. Apart from the long-wavelength behavior the appearance of $F_1^{(s)}(q)$ is very similar to $F_0^{(s)}(q)$ indicating that the zero-order static quantities $c_0(q)$ and $S_0(q)$ are also the relevant quantities for the higher order self-nonergodicity parameters, since the static quantities $c_\mu(q)$ and $S_\mu(q)$ are significantly different between $\mu=0$ and $\mu=1$. Due to the periodic boundary conditions and the absence of layering, differences between the separate modes are less significant compared to the slit case~\cite{Mandal:SoftMatter:2017}.

\subsection{SISF}
\begin{figure}[htp]
\begin{minipage}[b]{\linewidth}
\centering
\includegraphics{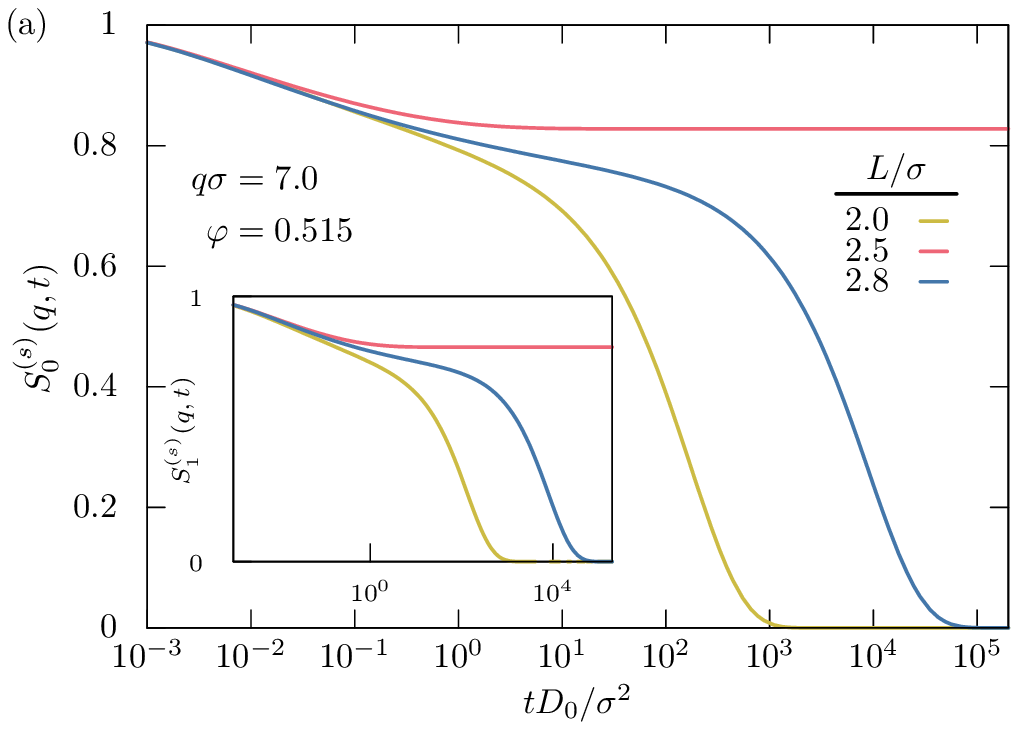}
\end{minipage}
\begin{minipage}[b]{\linewidth}
\centering
\includegraphics{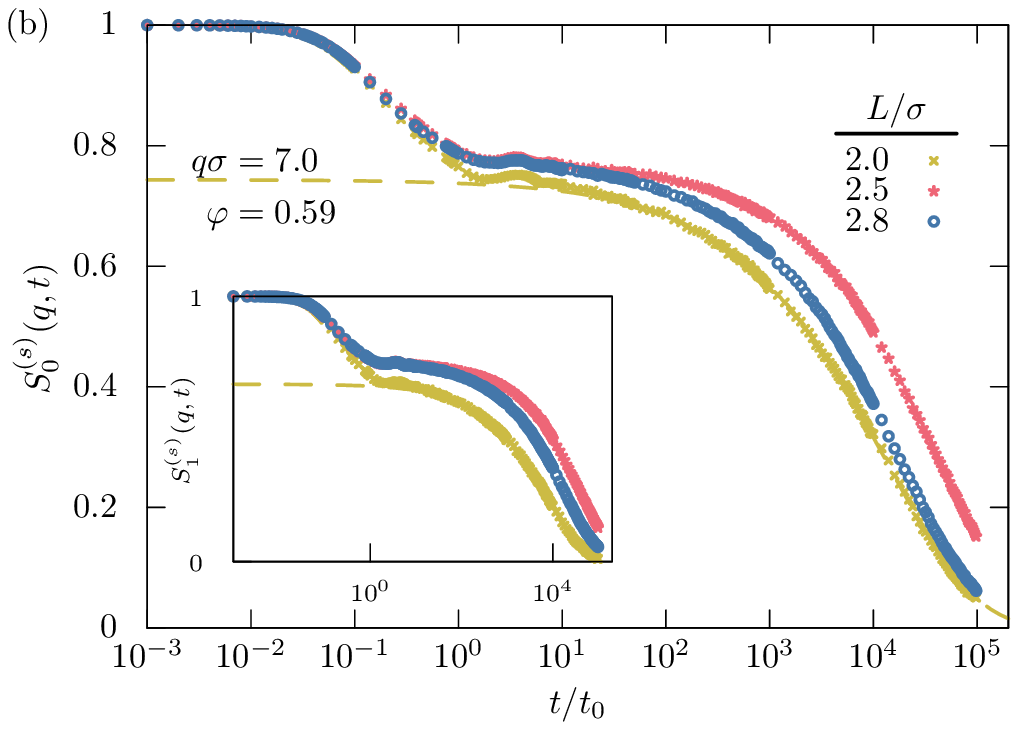}
\end{minipage}
\caption{SISF $S^{(s)}_0(q,t)$ and $S^{(s)}_1(q,t)$ (inset) for $q\sigma=7.0$ at (a) packing fraction $\varphi=0.515$ for MCT and (b) $\varphi=0.59$ for simulations, respectively. The dashed line in (b) represents a Kohlrausch-William-Watts stretched exponential fit to the simulation data. For clarity it is only shown for the smallest confinement length.}
\label{fig:SISF}
\end{figure}

We proceed with the full dynamic solution of the SISF. The unit of time for the MCT results is set by $\sigma^2/D_0$. \Fref{fig:SISF} (a) shows the first two modes $S_0^{(s)}(q,t)$ and  $S_1^{(s)}(q,t)$ (inset) for MCT at packing fraction $\varphi=0.515$, where it is known from the nonequilibrium-state diagram that a reentrant behavior occurs~\cite{Schrack:JStatMech:2020}. The related data for simulations at the higher packing fraction $\varphi=0.59$ are displayed in \fref{fig:SISF} (b). 

The MCT results look quite similar to the corresponding collective ISF~\cite{Schrack:JStatMech:2020} with clearly observable non-monotonic behavior depending on the confinement length. The MCT solution either decays to zero for the more commensurate values ($L=2.0\sigma$, $L=2.8\sigma$), identifying liquid states, or it reaches a non-vanishing plateau value corresponding to an arrested glassy state for the incommensurate value ($L=2.5\sigma$). The increased longitudinal diffusion in the case of commensurate packing prevents slowing down of the dynamics, and therefore, the arrested glassy state is shifted to higher packing fractions compared to incommensurate packing~\cite{Schrack:JStatMech:2020}. 

The behavior of the first higher-order mode $S^{(s)}_1(q,t)$ almost duplicates the one of $S^{(s)}_0(q,t)$. Whereas $S^{(s)}_0(q,t)$  only considers the dynamics parallel to the confining direction, higher order modes of the SISF also include the perpendicular direction. The similarity between the modes of the SISF suggests that not only the long-time limits, but the whole dynamics are mainly driven by the zero-order mode of the static quantities, $S_0(q)$ and $c_0(q)$, in particular by the  first sharp diffraction peak, like in the collective case~\cite{Schrack:JStatMech:2020}. Notably, $S_1^{(s)}(q,t)$ decays faster than $S_0^{(s)}(q,t)$ for all studied lengths.

\begin{figure}[htp]
\begin{minipage}[b]{\linewidth}
\centering
\includegraphics{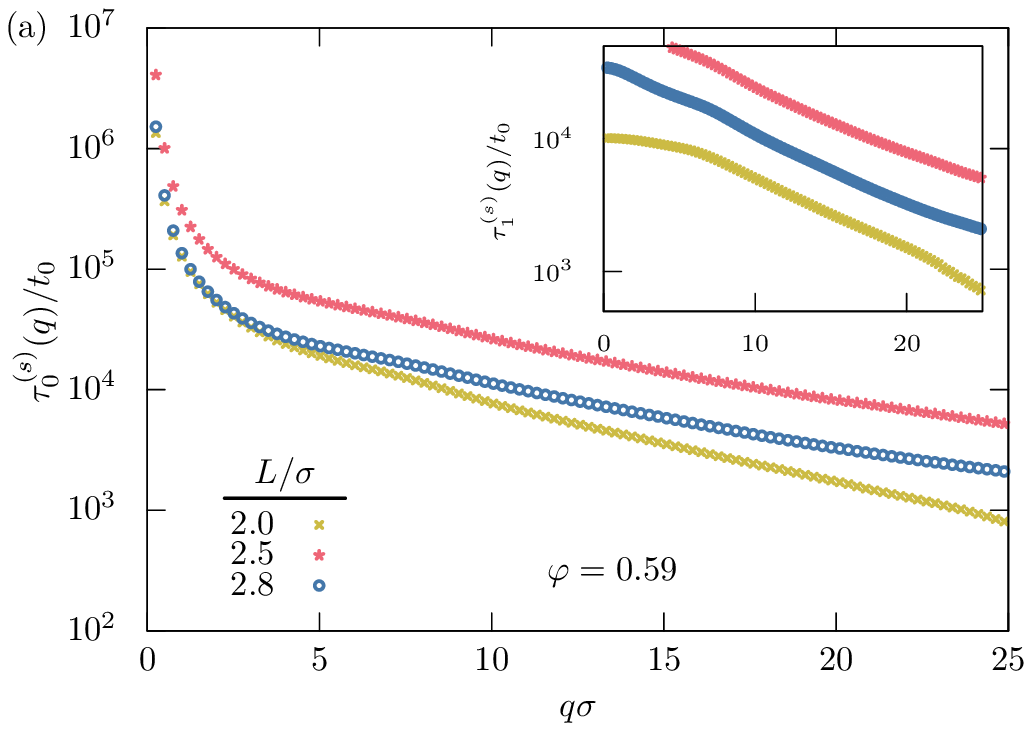}
\end{minipage}
\begin{minipage}[b]{\linewidth}
\centering
\includegraphics{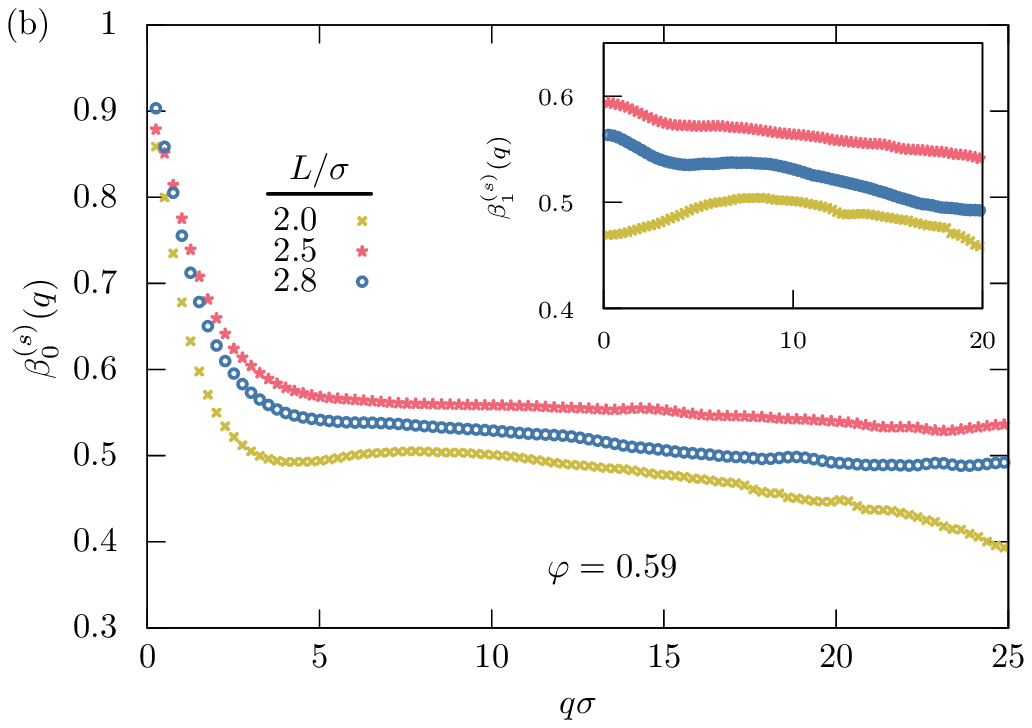}
\end{minipage}
\caption{Upper panel (a): self-relaxation times $\tau_0^{(s)}(q)$ and $\tau_1^{(s)}(q)$ (inset) for different values of $L$ at packing fraction $\varphi=0.59$ extracted from simulations. Lower panel (b): Kohlrausch exponents $\beta_0^{(s)}(q)$ and $\beta_1^{(s)}(q)$ (inset).}
\label{fig:tau}
\end{figure}

In close resemblance to the collective ISF, the simulation data only show an intermediate plateau for all studied lengths, accompanied by a stretched relaxation. The general shape of the SISF coincides with the collective ISF with non-monotonic dependence on the confinement length. Yet, there are minor differences compared to the ISF, in particular the precise plateau values (incoherent nonergodicity parameters) differ as it has been discussed in the preceding section. Even though the non-monotonic dependence is not obvious in the plateau value, it still can be observed in the corresponding wavenumber dependent relaxation times $\tau_0^{(s)}(q)$ (\fref{fig:tau} (a)) and the Kohlrausch exponents $\beta_0^{(s)}(q)$ (\fref{fig:tau} (b)). It would be interesting to analyze if the large-wavenumber limit of the Kohlrausch exponent coincides with the von Schweidler exponent similar to other recent studies comparing the von Schweidler and the KWW equation~\cite{Ruscher:JPCM:2020}. Unfortunately, we cannot accurately determine the von Schweidler exponent form our data. Even for higher order modes, where the SISF almost overlap (inset of \fref{fig:SISF} (b)), the related relaxation times $\tau_1^{(s)}(q)$ and Kohlrausch exponents $\beta_1^{(s)}(q)$ reveal considerable non-monotonic behavior (\fref{fig:tau} (insets)). Therefore, we conclude that within the simulations the non-monotonic dependence is absent to a large extend in the plateau heights, but it is still present in the dynamics manifested by the relaxation times and the Kohlrausch exponents. Simulation data also confirms that higher modes of the SISF decay faster than the zero order quantity.

\subsection{MSD}\label{sec:results_msd}
\begin{figure}[htp]
\begin{minipage}[b]{\linewidth}
\centering
\includegraphics{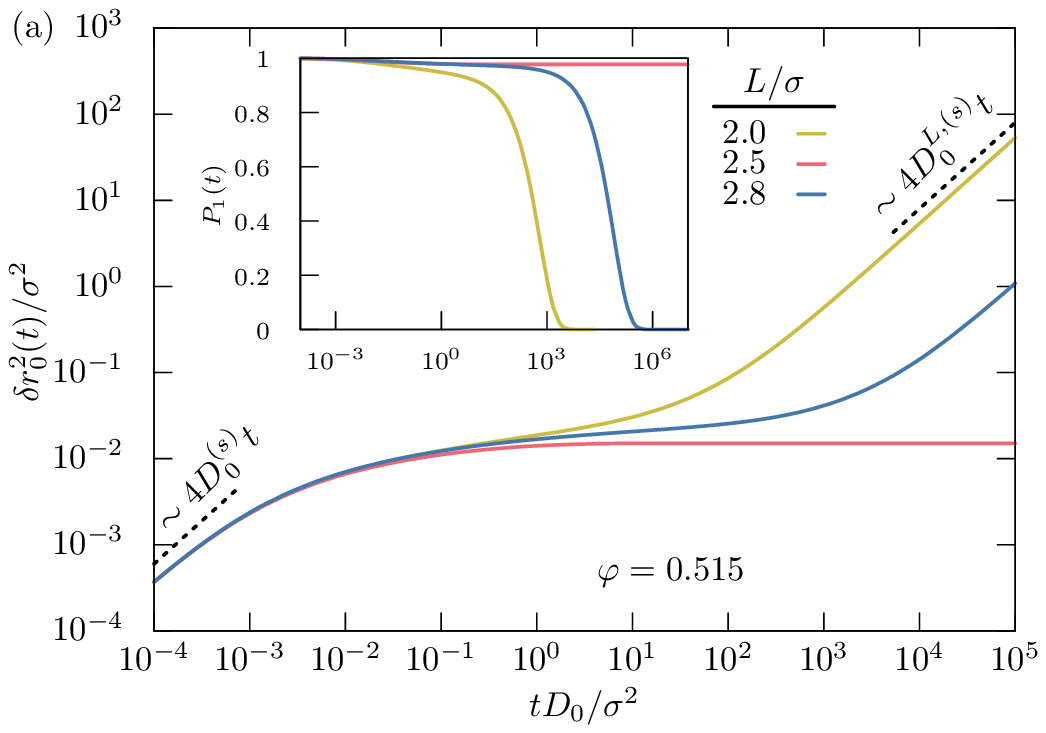}
\end{minipage}
\begin{minipage}[b]{\linewidth}
\centering
\includegraphics{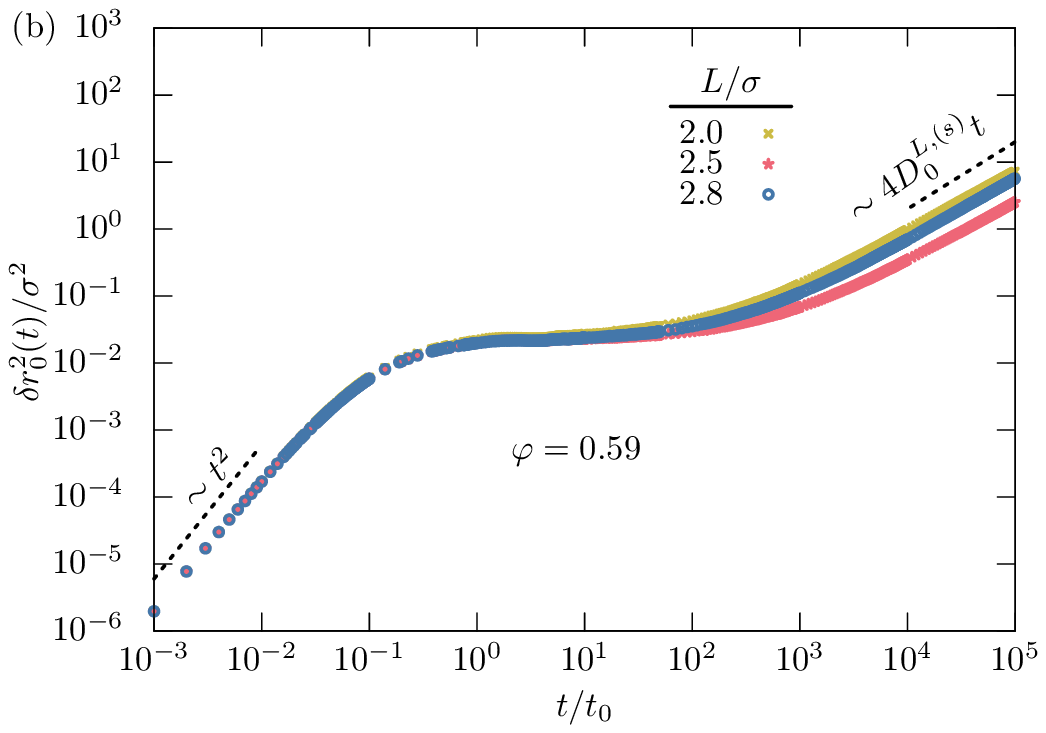}
\end{minipage}
\caption{MSD $\delta r_0^2(t)$ at (a) packing fraction $\varphi=0.515$ for MCT and (b) $\varphi=0.59$ for simulations. The inset in (a) shows $P_1(t)$.}
\label{fig:MSD}
\end{figure}

The mean-square displacement (MSD) is calculated for different confining lengths $L$ at the same packing fractions $\varphi=0.515$ for MCT (\fref{fig:MSD} (a)) and $\varphi=0.59$ for simulations (\fref{fig:MSD} (a)) as above for the SISF. 

Quantitative differences in the limit of small $t$ arise due to different microscopic dynamics. Whereas the MCT equations are derived for Brownian microscopic dynamics characterizing colloidal liquids, the simulations rely on Newtonian molecular dynamics. Therefore, for MCT, diffusive motion $\delta r^2_0(t)\simeq 4D_0 t$, $t\to 0$ can be observed for all investigated confinement lengths in the limit of small $t$, in contrast to the simulations, which show ballistic motion. With increasing time the dynamics are slowed down in the crossover regime as has already been discussed in connection with the SISF. For even larger times, particles are trapped due to the cage effect in the glassy state ($L=2.5\sigma$), which becomes apparent in the MCT solution by a plateau value of the MSD. In contrast, the data for the other confinement lengths show liquid-like behavior with long-time diffusion $\delta r^2_0(t)=4D_0^{L,(s)} t$, $t\to\infty$ after the crossover regime with a two-step relaxation process distinctive for MCT dynamics. This second rise is the $\alpha$-process of the MSD, indicating the tagged-particle's exit of the cage. For simulations, particles move out of the cage eventually for all confinement lengths, and diffusive motion can be observed at long times. This behavior is consistent with the intermediate nature of the plateau in the SISF.

The inset of \fref{fig:MSD} (a) shows the non-monotonic behavior of $P_1(t)$ within MCT. The general shape of the relaxation is quite similar compared to the SISF for small $q$. The curves are indistinguishable for small times $tD_0/\sigma^2 \lesssim 10^{-1}$ before the crossover regime sets in. All liquid curves decay to zero and the finite value for glassy states corresponds to $F^{(s)}_1(q\to 0)$ as already mentioned above.

A characteristic localization length $r_0^{(s)}$ can be defined by the small-$q$ behavior of the nonergodicity parameter, $F_0^{(s)}(q)=1-\left(q r_0^{(s)}\right)^2 + \mathcal{O}(q^4)$ similar to the bulk case~\cite{Fuchs:PRE:1998}, which is related to the long-time limit of $\delta r_0^2(t)$ by

\begin{eqnarray}
\lim_{t\to\infty} \delta {r}_0^{2} (t) = 4 r_0^{(s)2}.
\end{eqnarray}

In the glassy state ($L=2.5\sigma$ in \fref{fig:MSD}) the finite value of the localization length $r_0^{(s)2}=\lim_{t\to\infty} \delta {r}_0^{2} (t)/4$ related to the MSD corresponds to the size of cages in which the colloidal particles are trapped and reflects the behavior of colloidal suspensions in experiments for the slit geometry~\cite{Nugent:PRL:2007,Edmond:PRE:2012}, yet no physical boundaries are present in our quasi-confined system.

\Tref{tab:Stokes_Einstein} indicates that the product of the long-time self-diffusion coefficient $D_0^{L,(s)}$ and the self-relaxation time $\tau_0^{(s)}$ at the structure factor peak, $q\sigma=7.0$, is approximately constant, albeit the individual contributions vary significantly (by a factor of 50 for MCT and 3 for simulations respectively) with the confinement length. Within MCT, we identify the relaxation time $\tau_0^{(s)}$ with the time where the SISF has reduced to $1/e$. Since the dynamics are glassy within MCT for $L=2.5\sigma$ no self-relaxation time and self-diffusion coefficient can be extracted. MCT predicts a single $\alpha$-relaxation scale, which implies the validity of the Stokes-Einstein (SE) relation for straight paths approaching the glass-transition line. However, since a reentrant phenomenon occurs between $L=2.0\sigma$ and $L=2.8\sigma$, the SE relation does not have to hold here. Nevertheless, it is approximately fulfilled, albeit a slightly non-montonic behavior of $\tau_0^{(s)}D_0^{L,(s)}$ within the simulations. It is known that the violation of the SE relation sets in close to the critical packing fraction\cite{Kumar:JCP:2006, Biroli:JoP:2007, Sengupta:JCP:2013, Mandal:SoftMatter:2018}.

\begin{table}[htp]
\caption{\label{tab:Stokes_Einstein} Self-relaxation time $\tau_0^{(s)}$ for $q\sigma=7.0$, long-time self-diffusion coefficient $D_0^{L,(s)}$ and the product $\tau_0^{(s)}D_0^{L,(s)}$ for different values of $L$ at (a) packing fraction $\varphi=0.515$ for MCT and (b) $\varphi=0.59$ for simulations, respectively. For $L=2.5$ the system is in the glassy state within MCT.}

\begin{indented}
\lineup
\item[]\begin{tabular}{@{}*{5}{l}}
\br                              
&$L/\sigma$&$\tau_0^{(s)}/\sigma^{2}D_0^{-1}$&$D_0^{L,(s)}/D_0$&$\tau_0^{(s)}D_0^{L,(s)}/\sigma^2$\cr 
\mr
	&2.0&$1.11\cdot 10^{2}$&$1.41\cdot 10^{-4}$&$ 1.57\cdot 10^{-2}$\cr
(a) MCT	&2.5&---& --- & --- \cr
	&2.8&$5.33\cdot 10^{3}$&$2.62\cdot 10^{-6}$&$ 1.40\cdot 10^{-2}$\cr
\br
&$L/\sigma$&$\tau_0^{(s)}/t_0$&$D_0^{L,(s)}/\sigma^2t_0^{-1}$&$\tau_0^{(s)}D_0^{L,(s)}/\sigma^2$\cr 
\mr
			&2.0&$1.36\cdot 10^{4}$&$1.81\cdot 10^{-5}$&$ 2.46 \cdot 10^{-1} $\cr
(b) Simulations	&2.5&$4.14\cdot 10^{4}$&$6.38\cdot 10^{-6}$&$ 2.64 \cdot 10^{-1}$\cr
			&2.8&$1.76\cdot 10^{4}$&$1.43\cdot 10^{-5}$&$ 2.45 \cdot 10^{-1}$\cr
\br
\end{tabular}
\end{indented}
\end{table}

\subsection{VACF}\label{sec:results_vacf}
\begin{figure}[htp]
\centering
\includegraphics{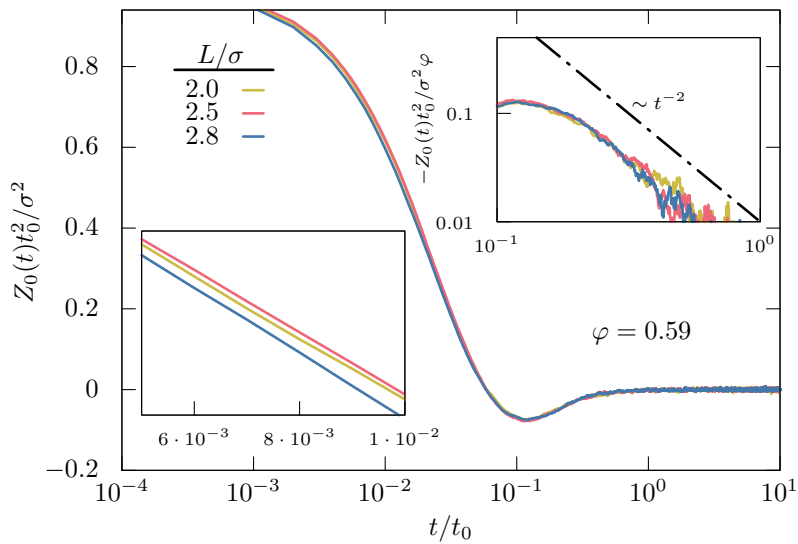}
\caption{Normalized VACF $Z_0(t)$ at $\varphi=0.59$ for simulations. The minimum indicates the caging of particles. Non-monotonic behavior is only weakly pronounced (lower left inset). The upper right inset shows the reduced VACF $Z_0(t)/\varphi$ depending on the confinement length $L$. The black dot-and-dash line labeled $t^{-2}$ serves as a guide to the eye.}
\label{fig:VACF_SIM}
\end{figure}

Due to the dominating linear growth of diffusion some subtleties are usually hidden in the MSD, therefore, it is instructive to also investigate the VACF. \Fref{fig:VACF_SIM} shows the VACF $Z_0(t)$ extracted from simulations for different confinement lengths. The prominent minimum in the VACF is related to the strong caging of particles for intermediate time scales. Quite similar curve shapes are present in the glass-like behavior of confined supercritical Argon~\cite{Ghosh:PRE_98:2018, Ghosh:SciRep:2019} with inhomogeneous density profiles.

As can be seen, non-monotonic behavior for short and intermediate times is only marginally pronounced, but still present (lower left inset). Since the crossover regime within the MSD for liquid states (\fref{fig:MSD}) extends over several orders of magnitude in time and just appears as an interpolation between the short-time diffusive motion $\delta r^2_0(t)=4D_0^{(s)} t, t\to 0$ and the long-time diffusion $\delta r^2_0(t)=4D_0^{L,(s)} t,t\to \infty$, we particularly want to investigate correlations of the VACF in the long-time regime. Unfortunately, due to insufficient statistics for the simulation data it is not possible to study long-time effects for the simulations, and we limit our discussion to MCT results. In principle, advanced simulation algorithms for colloidal bulk liquids have been developed recently to investigate the long-time behavior also for simulations~\cite{Mandal:PRL:2019}.

The numerical evaluation within MCT is not straightforward and we have to rely on a logarithmic wavenumber grid similar to the 3D bulk case~\cite{Mandal:PRL:2019} to extract the long-time behavior. The reduced wavenumbers are then given by $[x^{-N},\dots,x,1]q_{\mbox{\scriptsize{max}}}\sigma$ with high-$q$ cutoff $q_{\mbox{\scriptsize{max}}}\sigma = 40$, base $x=1.02$ and $N=400$ grid points. This corresponds to a minimal wavenumber $q_{\mbox{\scriptsize{min}}}\sigma=0.0148$, ensuring that the long-time behavior of the VACF is properly resolved. Additionally, only $|\mu|\le 10$ modes are used to save computational resources.

In principle, there are more sophisticated grid types which have been discussed lately for two- and three-dimensional systems~\cite{Caraglio:JCP:2020}. The main idea behind these nonuniform grids is to have a better representation with more grid points near the maxima and minima of the structure factor and the direct correlation function, but less wavenumbers in regions where the behavior is more linear. The implementation of this scheme for quasi-confined liquids would require a different wavenumber grid for each mode $\mu$, since the extreme values of the static input sensitively depend on the wavenumber  for different mode indices~\cite{Petersen:JStatMech:2019}. Still, we find for the long-time behavior of the VACF it is sufficient to have a good long-wavelength resolution since the corresponding integrals are dominated by low wavenumbers instead of the structure factor peaks, and therefore, we simply rely on a logarithmic grid independent of the mode index.
 
\begin{figure}[htp]
\centering
  \includegraphics{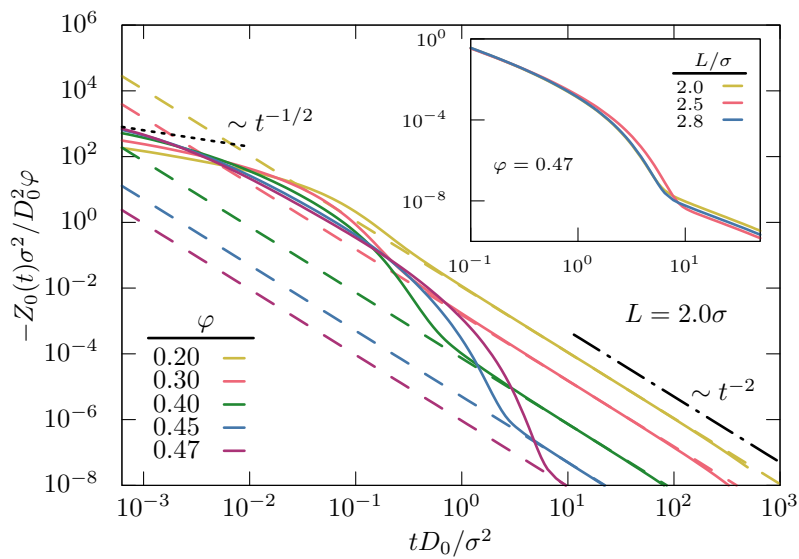}
  \caption{Reduced VACF $Z_0(t)/\varphi$ at constant confining length $L=2.0\sigma$ for increasing packing fraction $\varphi$ calculated from MCT. The black dotted line labeled $t^{-1/2}$ and the black dot-and-dash line labeled $t^{-2}$ serve as a guide to the eye and indicate the short-time and long-time behavior respectively. The colored dashed lines are the long-time asymptotes calculated from MCT according to equation~\eref{eq:VACF_LTT}. The inset shows the reduced VACF depending on the confinement length $L$ at packing fraction $\varphi=0.47$.}\label{fig:tail_VACF}
\end{figure}

At the arbitrarily chosen confinement length $L=2.0\sigma$ the numerical results for the reduced VACF $-Z_0(t)/\varphi$ for MCT display non-trivial correlations beyond the crossover regime of the MSD with anti-correlated long-time tails $Z_0(t)\sim -t^{-2}$ for all investigated packing fractions, see \fref{fig:tail_VACF}. Yet, in contrast to the bulk behavior, $Z(t)\sim -t^{-5/2}$, the exponent indicates two-dimensional dynamics with the prefactor sensitively depending on the density, equation~\eref{eq:VACF_LTT}. Surprisingly, the non-monotonic dependence on the confinement length only becomes apparent for larger packing fractions $\varphi\gtrsim 0.45$ approaching the glass-transition (inset). For this comparatively high packing fraction non-monotonic behavior depending on the confinement length can be observed in two respects. Firstly, the crossover regime is shifted to larger times for $L=2.5\sigma$ compared to the other confining lengths and secondly, the tails are more suppressed for incommensurate values. This demonstrates that evidence of glassy-like behavior can be observed even at $\varphi=0.47$, since these are the same values of $L$ which correspond to glassy states at higher packing fraction. Indeed, the strong $L^3$ dependence in the denominator of equation~\eref{eq:VACF_LTT} is compensated for by the $L$-dependent structural quantities $S_0(0)$, $c_0(q)$ and the long-time self-diffusion coefficient $D_0^{L,(s)}$ if the density is not too high.

For large confinement lengths ($L\gtrsim 5.0\sigma$) the dynamics of the particle are indistinguishable from bulk motion for short times with an intermediate window, where the VACF decays with a power-law, $Z(t)\sim -t^{-5/2}$. Only for longer times, a crossover from the unconfined dynamics occurs and the expected power law for quasi-confinement, $Z(t)\sim -t^{-2}$, is observed, similar to the tagged-particle within a slit~\cite{Jung:PRE_2:2020}. Extreme confinement, however, cannot be considered as the two dimensional limit case since due to periodic boundary conditions the hard spheres are still able to move perpendicular to the confining direction.

Although, the long-time behavior in the simulations is covered by noise, indications of a similar intermediate tail $Z(t)\sim -t^{-2}$ can be observed in the upper right inset of \fref{fig:MSD}. Indeed, for much longer times the classical hydrodynamic Alder tails~\cite{Alder:PRA:1970} are expected for a momentum-conserving system of Newtonian hard-spheres. A comparable intriguing crossover from positive correlations to anti-correlations has been reported for dense hard-sphere systems~\cite{Williams:PRL:2006}.

All MCT data show a divergent short-time behavior, $Z(t)\sim -t^{-1/2}$ as $t\to 0$, related to the hard-sphere potential. This divergence is not restricted to quasi-confined systems and can be observed in colloidal liquids with arbitrary dimension, e.g.\ in 3D bulk liquids~\cite{Mandal:PRL:2019}. The exponent does not depend on the confining geometry, contrary to the long-time tail.

MCT also gives a qualitative explanation for the strong variation of the long-time tail in terms of the static compressibility and slowing down of transport related to the 3D bulk case~\cite{Mandal:PRL:2019}. In accordance with equation~\eref{eq:VACF_LTT} and reference~\cite{Mandal:PRL:2019} the tail is suppressed by orders of magnitude with increasing $\varphi$. In the dilute limit the self-diffusion coefficient in the denominator can be ignored and the prefactor of the long-time tail shows a direct dependence on the long-wavelength limit of the static structure factor, $S_0(0)$. Approaching the glass transition the tails get even more suppressed due to the singularity of $D_0^{L,(s)}$. At intermediate times we then expect to observe two additional power laws, which are directly related to the critical beta decay~\cite{Jung:PRE_2:2020}.

\section{Summary and conclusions}\label{sec:conclusion}
In this work, we have investigated the tagged-particle dynamics of quasi-confined colloidal liquids by mode-coupling theory and event-driven simulations. In particular, we have elaborated self-nonergodicity parameters, the self-intermediate scattering function (SISF), the mean-square displacement (MSD) and the velocity-autocorrelation function (VACF). By adapting our stable numerical algorithm to the self-dynamics we have found qualitative agreement between MCT and simulations, showing distinctive non-monotonic behavior with the confinement length. Within the simulations, the non-monotonic behavior in the SISF and incoherent nonergodicity parameters is slightly less pronounced compared to the collective case. Nevertheless, it is still present, particularly in the corresponding relaxation times. Since $S_0(t)$ decays significantly more slowly than $S_1(t)$ we conclude that the slowing down of the dynamics is most pronounced perpendicular to the confining direction. The behavior of the MSD nicely captures these features, and in the MCT clearly illustrates the transition from particle localization in the glassy state to particle diffusion in the liquid state.

The non-monotonic behavior can be explained by an interplay between the local packing of the hard spheres and the confinement itself with an alternation between commensurate (sliding motion) and incommensurate packing (obstruction). Recent investigations between the local structure ordering and reentrance phenomena in a slit geometry suggest that local ordering changes from hexatic order in the commensurate case to local square order for incommensurate geometries~\cite{Ryan:MolPhy:2020}. A similar approach in computer simulations tries to disentangle confinement and layering effects by applying an external potential to the unconfined system resulting in a similar density modulation as in the slit~\cite{Saw:JoCP:2016}. Although the density profiles are quite similar in these two systems, they remain inhomogeneous in contrast to our quasi-confined liquid.

The glass-transition line extracted from the nonequilibrium-state diagram separates liquid-like from non-ergodic states~\cite{Schrack:JStatMech:2020}. Due to the smooth and continuous transition, we expect the generic $A_2$ fold bifurcation in Arnol'd's terminology~\cite{Arnold:catastrophe}, although in principle higher-order singularities, where the critical scaling behavior at the $A_3$ critical endpoint has been analyzed~\cite{Nandi:PRL:2014}, are also possible within quasi-confined systems. In particular, these singularities should be present for competing mechanisms similar to bulk systems such as binary mixtures~\cite{Voigtmann:EPL:2011}, attractive colloids~\cite{Dawson:PRE:2000, Sciortino:PRL:2003}, porous media~\cite{Krakoviack:PRL:2005, Krakoviack:PRE:2007} or randomly pinned systems~\cite{Cammarota:PNAS:2012, Cammarota:EPL:2013}.

To sum up, we have completed our recent study on quasi-confined liquids~\cite{Schrack:JStatMech:2020} with investigating the tagged-particle dynamics. Even in the absence of layering in the density profile, there are pronounced non-monotonic effects, e.g.\ within the MSD, caused by the confinement itself. Generally, our system behaves similarly to strongly confined colloids within a slit~\cite{Jung:PRE_102:2020}. The MCT equations of the quasi-confined liquid are reminiscent of the diagonal approximation for the slit geometry. Therefore, a detailed MCT analysis~\cite{Jung:JStatMech:2020,Jung:PRE_102:2020,Jung:PRE_2:2020} can easily be adopted to quasi-confined liquids. Yet, we want to emphasize that for quasi-confined liquids it is an exact representation rather than an approximation. Besides, results between quasi-confinement and slit geometry always differ significantly due to important differences in the static input for the MCT equations.

Generally, MCT and simulations coincide slightly better here in comparison to the slit geometry, e.g.\ manifested in the nonergodicity parameters (cf.\ reference~\cite{Mandal:SoftMatter:2017}). This appears to be due to the homogeneous density profile in quasi-confined systems. In contrast, the pronounced layering within the slit is distorted in simulations by polydispersity.

For the VACF we have mainly focussed on the long-time behavior identifying an anti-correlated algebraic decay for long times with an exponent characterizing the effective two-dimensional dynamics perpendicular to the confinement. These correlations are hidden in the MSD and can be rationalized within MCT. Future investigations could reveal similar tails within the non-Gaussian parameter or the Burnett coefficient~\cite{Ernst:JoSP:1984, Machta:JoSP:1984}.

Our results are not limited to hard-sphere systems since several studies show the dynamic equivalence between soft colloids and hard sphere systems, e.g.\ recent experiments for soft star polymers~\cite{Wang:ACSMacro:2019}. Non-monotonic confinement effects should also be present in driven systems like granular matter~\cite{Kranz:PRL:2010, Sperl:EPL:2012, Kranz:PRE:2013} and active microrheology~\cite{Gruber:PRE:2016,Gruber:PRE:2020}.

\ack
This work has been supported by the Austrian Science Fund (FWF): I 2887. CFP gratefully acknowledges a Lise-Meitner fellowship of the Austrian Science Fund (FWF): M 2471. The computational results presented have been achieved in part using the HPC infrastructure LEO of the University of Innsbruck.

%\appendix
\section*{References}
\bibliography{dynamics_quasi_confinement_tagged}

\end{document}